\documentclass[a4paper,12pt]{article}

\usepackage{url}
\usepackage{epsfig}
\usepackage{amsmath}
\usepackage{amssymb}
\usepackage{amsfonts}
\usepackage{bbm}
\usepackage[footnotesize]{caption}
\usepackage{graphicx}
\usepackage{mathrsfs}
\usepackage[font=scriptsize]{subcaption}
\usepackage{color}
\usepackage[dvipsnames]{xcolor}
\usepackage{comment}
\usepackage{mathtools}
\usepackage{braket}
\usepackage{cite}
\usepackage{hyperref}
\usepackage{multirow}
\usepackage{relsize}
\usepackage{fullpage}
\usepackage{makecell}
\usepackage{enumitem,xcolor}
\usepackage{doi}

\begin{document}

\begin{titlepage}
\begin{center}
{\bf\Large   Complete Unification in 10d $E_8$} \\[12mm]
Alfredo Aranda$^{\ddag}$%
\footnote{E-mail: \texttt{fefo@ucol.mx}},
Francisco~J.~de~Anda$^{\star}$%
\footnote{E-mail: \texttt{fran@tepaits.mx}},
\\[-2mm]

\end{center}
\vspace*{0.50cm}
\centerline{$^{\ddag}$ \it
Facultad de Ciencias-CUICBAS, Universidad de Colima, C.P.28045, Colima, M\'exico 01000, M\'exico}
\centerline{Dual CP Institute of High Energy Physics, C.P. 28045, Colima, M\'exico}
\vspace*{0.2cm}
\centerline{$^{\star}$ \it
Tepatitl{\'a}n's Institute for Theoretical Studies, C.P. 47600, Jalisco, M{\'e}xico}
\centerline{Dual CP Institute of High Energy Physics, C.P. 28045, Colima, M\'exico}
\vspace*{1.20cm}

\begin{abstract} 
A grand unification scenario is presented that is based on quantum field theory, and where a single $E_8$ gauge superfield in 10 dimensions is used to obtain all the particle content of the Standard Model at low energies. The key feature of the formulation lies in the dimensional reduction used to break the gauge symmetry and to determine the low energy spectrum. It is shown that, through the orbifold $T^6/(\mathbb{Z}_6\times \mathbb{Z}_2)$, and its corresponding Wilson lines, the symmetry is broken to the Standard Model one, generating a particular model that includes the Minimal Supersymmetric Standard Model spectrum. Furthermore it is also shown that the model is free of gauge anomalies at all levels by itself, i.e. without the need to include any additional representations of fields. Thus a complete unification of the Standard Model into a single gauge superfield is shown to be formally plausible. Although this letter does not include a phenomenological study of the specific model (currently being investigated), some interesting questions and observations are included as motivation for the scenario.
\end{abstract}
\end{titlepage}

\section*{Introduction}
The unification of all known gauge interactions into (perhaps) a single one has been a driving force for theoretical high energy physics for at least the past half century. During that time, several developments have enriched theoretical physics not only with specific models and plausible scenarios, but also with new techniques and methods that can and have been used in contexts other than unification.

The basic idea of gauge unification is based on the premise that there is a {\it large} gauge group that contains the Standard Model (SM) one, namely $SU(3)_C \times SU(2)_L \times U(1)_Y$, and that all other fields present in the SM are included in some of its representations. Then, upon symmetry breaking, one  obtains the exact structure of the SM at low energies, including the possibility of additional low energy states that might lead to interesting phenomenology~\cite{ Georgi:1974sy, Fritzsch:1974nn, Mohapatra:1974gc, Pati:1974yy}. 

The groups $SU(5)$, $SO(10)$ and $E_6$ have become prototypical gauge groups for Grand Unified Theories (GUTs), and models have been created by using them in many different scenarios. The literature is quite vast and robust, and the study of GUTs (including of course other gauge groups than the aforementioned) is a subject that has had a very prolific life (see~\cite{Langacker:1980js,Tanabashi:2018oca} for general description and general references). Among the most salient results and characteristics of GUTs one can mention the following: in part due to the nice feature of gauge coupling unification, most GUTs tend to be supersymmetric. This is certainly not a requirement, however supersymmetry (SUSY) does provide  several theoretical tools that make it appealing. Most GUTs predict proton decay and this puts a stringent constraint on the models (already excluding the simplest ones) that must be considered.

Although GUTs are constructed and analyzed within the realm of quantum field theory (QFT), string theory based constructions have been widely explored as well. This has led to a very active sub-field of the so-called string phenomenology that attempts to connect string theory results with high energy (low energy) physics. In particular, the role of geometry (compactification and orbifolding) has been an important ingredient in the effort to generate plausible GUTs from string theory based scenarios.  

An interesting and welcome consequence of these efforts has been the use of extra dimensions in the context of QFT that has also allowed the formulation of non-stringy, extra dimensional arguments and modeling, providing more routes of exploration \cite{Nilles:2014owa}.

In this letter, $E_8$ is used as the gauge symmetry of a unification model in $10$ dimensions~\footnote{The exploration of these ideas is quite old and interesting. In fact, it is noteworthy that once the anomaly cancellation mechanism in string theory appeared in the mid-eighties, seeding the so-called second string revolution, the attention strongly shifted towards string-related scenarios and some of the {\it old-fashioned} ideas were put on hold. The mechanism proposed in this letter attempts to bridge some of them back. }. Among the gauge groups that one can use to formulate a GUT, $E_8$ has one particular {\it virtue} that makes it very attractive: its adjoint representation ($\textbf{248} $) is also the fundamental representation. This allows the complete particle content of the SM (fermions and bosons) to lie in the same representation \cite{Slansky:1981yr}. Of course, one then needs to find a way to dismantle this representation into what is observed at low energies. Note that the fact that it is possible to include {\it everything} into the adjoint representation immediately suggests an $\mathcal{N}=4$ super Yang Mills (SYM) theory in four dimensions (4d), which has the property of being completely finite. While this is nice in principle, it is very far from a realistic scenario. In fact, $E_8$ is commonly "passed on" due to the fact that it only has real representations and therefore mechanisms to generate chirality must be introduced~\cite{Adler:2002yg}. An interesting mechanism is found in scenarios with extra dimensional orbifolds. Using the framework of orbifolding, it has been recently shown that one can consistently unify all the SM gauge fields, fermions and scalars, into a single $E_8$ gauge superfield in 10 dimensions, that upon orbifold breaking generates the Minimal Supersymmetric Standard Model (MSSM) and a flavor gauge symmetry. This can be accomplished with the orbifold $T^6/(\mathbb{Z}_6\times \mathbb{Z}_2)$~\cite{Aranda:2020noz}. This model however requires the introduction of several additional fields in order to cancel anomalies associated to the remaining flavor symmetry. This is done by suitably  adding certain appropriate representations of fermions localized in specific branes to cancel the anomalies, as is typically done in string motivated constructions.

The logic behind the construction presented in~\cite{Aranda:2020noz} resides in the fact that one can start from $\mathcal{N}=1\ \ E_8$ SYM in 10d, compactified using an orbifold which breaks both the $E_8$ and the would-be extended $\mathcal{N}=4$ SUSY in 4d, to the SM group with $\mathcal{N}=1$ SUSY in 4d, while removing the mirror fermions (included in $E_8$ models to solve the chirality problem)~\cite{Olive:1982ai,Babu:2002ti}. 

These ideas appear naturally in superstring theory, however they are independent of strings and rarely discussed in field theory. This motivates the following question: Is it possible, within the QFT framework alone, to find the appropriate geometry (namely orbifold) for an $E_8$ model with a single adjoint representation, that renders the SM particle content and gauge group at low energies without any additional fields to cancel anomalies? i.e. Is it possible to have a pure gauge - {\it complete} - $E_8$ unification model consistent with the SM (particle content and gauge group)?

In this letter an affirmative answer to this question is given. A consistent model is presented, with a single $E_8$ gauge superfield in 10d, broken through orbifolding into the MSSM. Furthermore it is shown that the theory is anomaly free at all levels. This does not happen in usual $E_8$ based string models, as several $(\textbf{248})$ representations are needed to reproduce the SM, as well as localized states to cancel the anomalies. Thus, the message intended for this communication can be succinctly stated as follows:  It is possible to build a SM-consistent GUT consisting only of the vector superfield associated to $E_8$ in $10$d, without the need of any extra field content \footnote{Please note that the purpose of this letter is to show a proof of concept and that a fully realistic model requires a comprehensive phenomenological study. Such study is being carried out and will be presented in a separate publication.
}.

To accomplish it, the scenario involves both symmetry breaking by the orbifold and the Wilson lines associated to it, so it is completely a geometrical construction: the orbifolding breaking uses boundary conditions on rotations of the extra dimensional coordinates while Wilson lines use boundary conditions on translations; both effectively provide  different masses to different states.  These {\it simultaneous} boundary conditions define the orbifold through a non trivial twist of the extra dimensional components of the Poincar\`e group. 

The key observation is that the low energy effective theory is a result of the integration of massive Kaluza-Klein (KK) modes as well as some would-be zero modes that obtain mass through Wilson lines (the mass scale is the same for both, namely the compactification scale). Thus, the low energy theory is composed of the remaining modes and, importantly, it must be anomaly free by itself. As described below, this is non trivial and strongly restricts the possibilities, leading to a very definite model .

\section{Super Yang-Mills in $T^6/(\mathbb{Z}_6\times \mathbb{Z}_2)$}
The theory is an $\mathcal{N}=1$ SYM theory in $10$d for the gauge group $E_8$.
There is a single $10$d vector gauge superfield $\mathcal{V}(x,z_1,z_2,z_3)$ (where $x$ denotes the uncompactified 4d coordinates in $R^4$ 
and the $z_i$ denote three complex coordinates of the remaining
compact $6$d space) that decomposes into a $10$d vector field and a $10$d Weyl/Majorana fermion. This superfield is in the $\mathcal{V}_{(248)}\sim  \textbf{248}$ adjoint representation of $E_8$ (which coincides with the fundamental representation). The $10$d vector superfield $\mathcal{V}$ decomposes into a $4$d vector superfield $V$ and three $4$d chiral superfield multiplets $\phi_{1,2,3}$, which implies $\mathcal{N}=4$ SUSY in 4d after compactification, with a lagrangian \cite{ArkaniHamed:2001tb,Brink:1976bc}.
\begin{eqnarray} 
\mathcal{L} =\frac{1}{32}  \tau\int d^2\theta\  W^\alpha W_\alpha+\int d^2\theta d^2\bar{\theta}\bar{\phi}^i e^{2gV}\phi^i- 
 \int d^2\theta\sqrt{2} g \phi_1[\phi_2,\phi_3]+h.c. ,
\label{eq:n41lag}
\end{eqnarray}
where $i= 1,2,3$. Note the explicit $SU(3)_R\times U(1)_R$ symmetry remaining from the rotation between the three complex coordinates and the complex rotation in all of them. The simple SUSY related to $U(1)_R$ is separated from the extended one, related to $SU(3)_R$. 

The extra dimensions have the Poincar\`e symmetry $O(6)\ltimes T^6/\Gamma$, where $O(6)$ corresponds to rotations, and $T^6$ corresponds to translations. The translation group is modded by the lattice vectors $\Gamma=\{\tau^a_i\}\simeq\mathbb{Z}^6$ which makes the space compact $\mathbb{R}^6\to T^6\simeq\mathbb{R}^6/\Gamma$ due to the periodicity conditions, $z_i\sim z_i +\tau_i^r$, where $i=1,2,3,$ for each complex dimension and $r=1,2,$ for each basis vector for a torus. For the $T^6/(\mathbb{Z}_6\times \mathbb{Z}_2)$ torus these are defined as 
\begin{equation}
\begin{split}
\tau_i^1=1,
\ \ \ (\tau_1^2,\tau_2^2,\tau_3^2)=\left(e^{i\pi/3},\frac{e^{i\pi/6}}{\sqrt{3}},\frac{e^{i\pi/6}}{\sqrt{3}}\right).
\end{split}
\end{equation}
The extra dimensions are further assumed to be orbifolded by a discrete group $F$ so that the actual extra dimensional space is $T^6/F$ where $F\in O(6)$ is a discrete subgroup of the rotation group. The group $F$ must be a symmetry of the lattice $F\Gamma=\Gamma$ to consistently define an orbifold. The rotation group is $O(6)\simeq SO(6)\times \mathbb{Z}_2\simeq SU(4)\times \mathbb{Z}_2$. 
 If one desires to keep simple SUSY after orbifolding (leave an unbroken $U(1)_R$), then $F\subset SU(3)$. The most general Abelian orbifolding that preserves $\mathcal{N}=1 $ SUSY is 
$F\simeq \mathbb{Z}_N \times  \mathbb{Z}_M\subset SU(3)$,
with positive integers $N,M$ \cite{Fischer:2012qj}.
From them, the $T^6/(\mathbb{Z}_6\times \mathbb{Z}_2)$ is singled out as the one closest to the SM \cite{Aranda:2020noz}.

The $T^6/(\mathbb{Z}_6\times \mathbb{Z}_2)$ orbifold is defined as:
\begin{equation}\begin{split}
(x,z_1,z_2,z_3) &\sim (x,\alpha^2 z_1,\alpha^5 z_2, \alpha^5 z_3),\\
(x,z_1,z_2,z_3) &\sim (x, -z_1,- z_2, (-1)^2 z_3),
\label{eq:orbrot}
\end{split}\end{equation}
where $\alpha=e^{2i\pi/6}$ and $-1=e^{2i\pi/2}$ . 
The $\mathbb{Z}_6$ and $\mathbb{Z}_2$ rotations are consistent as they are equivalent to lattice transformations. 

 There are no non-trivial discrete Wilson lines in a $\mathbb{Z}_6$ orbifold. There can be three independent continuous Wilson lines which have to be aligned with the representations that have zero modes, to be discussed below. It is important to emphasize that these Wilson lines are associated to the existence of the orbifold. They are defined by the boundary conditions on ED translations while the orbifolding is defined by boundary conditions on ED rotations, so they should be treated on the same footing.
 
 \section{$E_8$ to SM through orbifolding}

The choice of the orbifold is
\begin{equation}
\mathbb{Z}_6: \phi \to e^{2i\pi q_{X'}/6}\phi,\   \mathbb{Z}_2: \phi \to e^{2i\pi (q_{Y}+q_{T_F^8})/2}\phi,
\end{equation}
which breaks $E_8\to SU(4)_{PS}\times SU(2)_L\times SU(2)_R\times U(1)_{X'}\times U(1)_F\times SU(2)_F$. 

The 10d gauge superfield $\mathcal{V}_{(\textbf{248})}$ decomposes as in table \ref{tab:pps2} with their corresponding orbifold charges (color coding is \textcolor{magenta}{adjoints}, \textcolor{blue}{SM fermions}, \textcolor{green}{Higgs bosons}, \textcolor{violet}{right handed neutrinos}, \textcolor{brown}{flavons}, \textcolor{red}{mirror fermions} and \textcolor{orange}{mirror Higgses}.).

\begin{table}
	\centering
	\scriptsize
	\renewcommand{\arraystretch}{1.1}
	\begin{tabular}[t]{l|llll}
		\hline
		 & $V$ & $\phi_1$ & $\phi_2$ & $\phi_3$\\ 
		\hline
	$\mathcal{V}_{\textcolor{magenta}{(\textbf{15},\textbf{1},\textbf{1},0,0,\textbf{1})}} $ & $1,1$ & $\alpha^2, -1$ & $\alpha^5, -1$& $\alpha^5, 1$\\
		$\mathcal{V}_{\textcolor{magenta}{(\textbf{1},\textbf{3},\textbf{1},0,0,\textbf{1})}} $ & $1,1$ & $\alpha^2, -1$ & $\alpha^5, -1$& $\alpha^5, 1$\\
		$\mathcal{V}_{\textcolor{magenta}{(\textbf{1},\textbf{1},\textbf{3},0,0,\textbf{1})}} $ & $1,1$ & $\alpha^2, -1$ & $\alpha^5, -1$& $\alpha^5, 1$\\
		$\mathcal{V}_{\textcolor{magenta}{(\textbf{1},\textbf{1},\textbf{1},0,0,\textbf{1})}} $ & $1,1$ & $\alpha^2, -1$ & $\alpha^5, -1$& $\alpha^5, 1$\\
		$\mathcal{V}_{\textcolor{magenta}{(\textbf{1},\textbf{1},\textbf{1},0,0,\textbf{3})}} $ & $1,1$ & $\alpha^2, -1$ & $\alpha^5, -1$& $\alpha^5, 1$\\
		$\mathcal{V}_{\textcolor{magenta}{(\textbf{1},\textbf{1},\textbf{1},0,0,\textbf{1})}} $ & $1,1$ & $\alpha^2, -1$ & $\alpha^5, -1$& $\alpha^5, 1$\\
		$\mathcal{V}_{\textcolor{brown}{(\textbf{1},\textbf{1},\textbf{1},0,-3,\textbf{2})}} $ & $1,-1$ & $\alpha^2, 1$ & $\alpha^5, 1$& $\alpha^5, -1$\\
		$\mathcal{V}_{\textcolor{brown}{(\textbf{1},\textbf{1},\textbf{1},0,3,\textbf{2})}} $ & $1,-1$ & $\alpha^2, 1$ & $\alpha^5, 1$& $\alpha^5, -1$\\
		$\mathcal{V}_{(\textbf{6},\textbf{2},\textbf{2},0,0,\textbf{1})} $& $1,-1$ & $\alpha^2, 1$ & $\alpha^5, 1$& $\alpha^5, -1$\\
		$\mathcal{V}_{(\textbf{4},\textbf{2},\textbf{1},-3,0,\textbf{1})} $  & $\alpha^3,-1$ & $\alpha^5, 1$ & $\alpha^2, 1$& $\alpha^2, -1$\\
		$\mathcal{V}_{(\bar{\textbf{4}},\textbf{1},\textbf{2},-3,0,\textbf{1})} $ & $\alpha^3,1$ & $\alpha^5, -1$ & $\alpha^2, -1$& $\alpha^2, 1$\\
		$\mathcal{V}_{(\bar{\textbf{4}},\textbf{2},\textbf{1},3,0,\textbf{1})} $  & $\alpha^3,-1$ & $\alpha^5, 1$ & $\alpha^2, 1$& $\alpha^2, -1$\\		
		$\mathcal{V}_{(\textbf{4},\textbf{1},\textbf{2},3,0,\textbf{1})} $ & $\alpha^3,1$ & $\alpha^5, -1$ & $\alpha^2, -1$& $\alpha^2, 1$ \\	
		$\mathcal{V}_{(\textbf{6},\textbf{1},\textbf{1},-2,1,\textbf{2})} $ & $\alpha^4,-1$ & $1, 1$ & $\alpha^3,1$& $\alpha^3, -1$\\
		$\mathcal{V}_{(\textbf{6},\textbf{1},\textbf{1},-2,-2,\textbf{1})} $ & $\alpha^4,1$ & $1, -1$ & $\alpha^3,- 1$& $\alpha^3, 1$\\
		$\mathcal{V}_{(\textbf{6},\textbf{1},\textbf{1},2,-1,\textbf{2})} $  & $\alpha^2,-1$ & $\alpha^4, 1$ & $\alpha,1$& $\alpha, -1$\\	
		$\mathcal{V}_{(\textbf{6},\textbf{1},\textbf{1},2,2,\textbf{1})} $  & $\alpha^2,1$ & $\alpha^4, -1$ & $\alpha,- 1$& $\alpha, 1$\\	
		\hline
	\end{tabular}
	\hspace*{0.3cm}
	\begin{tabular}[t]{l|llll}
		\hline
		 & $V$ & $\phi_1$ & $\phi_2$ & $\phi_3$\\ 
		\hline
	$\mathcal{V}_{\textcolor{blue}{(\textbf{4},\textbf{2},\textbf{1},1,1,\textbf{2})}} $ & $\alpha,
	1$ & $\alpha^3, -1$ & $1, -1$& $1, 1$\\
	$\mathcal{V}_{\textcolor{blue}{(\textbf{4},\textbf{2},\textbf{1},1,-2,\textbf{1})}} $ & $\alpha,
	-1$ & $\alpha^3, 1$ & $1, 1$& $1, -1$\\
		$\mathcal{V}_{\textcolor{blue}{(\bar{\textbf{4}},\textbf{1},\textbf{2},1,1,\textbf{2})}} $ & $\alpha,-1$ & $\alpha^3, 1$ & $1, 1$& $1, -1$\\
		$\mathcal{V}_{\textcolor{blue}{(\bar{\textbf{4}},\textbf{1},\textbf{2},1,-2,\textbf{1})}} $ & $\alpha,1$ & $\alpha^3, -1$ & $1, -1$& $1, 1$\\
		$\mathcal{V}_{\textcolor{green}{(\textbf{1},\textbf{2},\textbf{2},-2,1,\textbf{2})}} $  & $\alpha^4,1$ & $1, -1$ & $\alpha^3, -1$& $\alpha^3, 1$\\
		$\mathcal{V}_{\textcolor{green}{(\textbf{1},\textbf{2},\textbf{2},-2,-2,\textbf{1})}} $  & $\alpha^4,-1$ & $1, 1$ & $\alpha^3, 1$& $\alpha^3, -1$\\
		$\mathcal{V}_{\textcolor{brown}{(\textbf{1},\textbf{1},\textbf{1},4,1,\textbf{2})}} $ & $\alpha^4,-1$ & $1, 1$ & $\alpha^3, 1$& $\alpha^3, -1$\\
		$\mathcal{V}_{\textcolor{brown}{(\textbf{1},\textbf{1},\textbf{1},4,-2,\textbf{1})}} $ & $\alpha^4,1$ & $1, -1$ & $\alpha^3, -1$& $\alpha^3, 1$\\
		$\mathcal{V}_{\textcolor{red}{(\bar{\textbf{4}},\textbf{2},\textbf{1},-1,-1,\textbf{2})}} $& $\alpha^5,1$ & $\alpha,-1$ & $\alpha^4, -1$& $\alpha^4, 1$\\
		$\mathcal{V}_{\textcolor{red}{(\bar{\textbf{4}},\textbf{2},\textbf{1},-1,2,\textbf{1})}} $& $\alpha^5,-1$ & $\alpha,1$ & $\alpha^4, 1$& $\alpha^4, -1$\\
		$\mathcal{V}_{\textcolor{red}{(\textbf{4},\textbf{1},\textbf{2},-1,-1,\textbf{2}) }} $  & $\alpha^5,-1$ & $\alpha, 1$ & $\alpha^4, 1$& $\alpha^4, -1$\\
		$\mathcal{V}_{\textcolor{red}{(\textbf{4},\textbf{1},\textbf{2},-1,2,\textbf{1})} } $  & $\alpha^5,1$ & $\alpha, -1$ & $\alpha^4, -1$& $\alpha^4, 1$\\
		$\mathcal{V}_{\textcolor{orange}{(\textbf{1},\textbf{2},\textbf{2},2,-1,\textbf{2})}} $ & $\alpha^2,1$ & $\alpha^4, -1$ & $\alpha, -1$& $\alpha, 1$\\
		$\mathcal{V}_{\textcolor{orange}{(\textbf{1},\textbf{2},\textbf{2},2,2,\textbf{1})}} $ & $\alpha^2,-1$ & $\alpha^4, 1$ & $\alpha, 1$& $\alpha, -1$\\	
		$\mathcal{V}_{\textcolor{brown}{(\textbf{1},\textbf{1},\textbf{1},-4,-1,\textbf{2})}} $  & $\alpha^2,-1$ & $\alpha^4, 1$ & $\alpha, 1$& $\alpha, -1$\\	
		$\mathcal{V}_{\textcolor{brown}{(\textbf{1},\textbf{1},\textbf{1},-4,2,\textbf{1})}} $  & $\alpha^2,1$ & $\alpha^4, -1$ & $\alpha, -1$& $\alpha, 1$\\	
		\hline
	\end{tabular}
	\caption{$\mathbb{Z}_6\times \mathbb{Z}_2$ orbifold charges of each $SU(4)_{PS}\times SU(2)_L\times SU(2)_R\times U(1)_{X'}\times U(1)_F\times SU(2)_F$ $\mathcal{N}=1$ superfield. Only the superfields with both charges equal to unity 
	(the singlets $1,1$) have zero modes.} 
	\label{tab:pps2}
\end{table}

The rotation boundary conditions give a different mass to each field with different orbifold charges in table \ref{tab:pps2}.  The zero modes are the ones in table \ref{tab:pps2} with (1,1) charge, namely
\begin{equation} 
\begin{split}
V_\mu &: \textcolor{magenta}{(\textbf{15},\textbf{1},\textbf{1},0,0,\textbf{1})}+\textcolor{magenta}{(\textbf{1},\textbf{3},\textbf{1},0,0,\textbf{1})}+\textcolor{magenta}{(\textbf{1},\textbf{1},\textbf{3},0,0,\textbf{1})}\\
&\quad+\textcolor{magenta}{(\textbf{1},\textbf{1},\textbf{1},0,0,\textbf{1})}+\textcolor{magenta}{(\textbf{1},\textbf{1},\textbf{1},0,0,\textbf{1})}+\textcolor{magenta}{(\textbf{1},\textbf{1},\textbf{1},0,0,\textbf{3})},\\
\phi_1&: (\textbf{6},\textbf{1},\textbf{1},-2,1,\textbf{2})+\textcolor{green}{(\textbf{1},\textbf{2},\textbf{2},-2,-2,\textbf{1})}+\textcolor{brown}{(\textbf{1},\textbf{1},\textbf{1},4,1,\textbf{2})},\\
\phi_2&: \textcolor{blue}{(\textbf{4},\textbf{2},\textbf{1},1,-2,\textbf{1})}+\textcolor{blue}{(\bar{\textbf{4}},\textbf{1},\textbf{2},1,1,\textbf{2})},\\
\phi_3&: \textcolor{blue}{(\textbf{4},\textbf{2},\textbf{1},1,1,\textbf{2})}+\textcolor{blue}{(\bar{\textbf{4}},\textbf{1},\textbf{2},1,-2,\textbf{1})},
\label{eq:zmf}
\end{split}\end{equation}
which can be named as
\begin{eqnarray} \nonumber
V_\mu &:& \textcolor{magenta}{G_\mu}+\textcolor{magenta}{W^L_\mu}+\textcolor{magenta}{W^R_\mu}+\textcolor{magenta}{Z'_\mu}+\textcolor{magenta}{Z^F_\mu}+\textcolor{magenta}{W^F_\mu},\\ \nonumber
\phi_1&:& T+\textcolor{green}{h}+\textcolor{brown}{\Phi},\ \ \ \
\phi_2: \textcolor{blue}{f}+\textcolor{blue}{F^c},\ \ \ \
\phi_3:\textcolor{blue}{F}+\textcolor{blue}{f^c},
\end{eqnarray}
where lower case letters correspond to $SU(2)_F$ singlets, corresponding to the first family while upper case letters are doublets corresponding to the second and third families. Note that this field content describes what the purely rotational boundary conditions do. It is presented only to show the symmetry breaking pattern and does not represent the physical spectrum at any scale, thus there are no anomalies associated to it.

 Some of the fields above are actually massive (with compactification scale mass) due to the Wilson lines. The Wilson lines also generate a mass splitting but in a different way. This is because the orbifolding adds boundary conditions to rotations, which form a compact group $SO(6)$, while the Wilson lines add boundary conditions to translations, which form a non compact group $T^6\simeq \mathbb{R}^6$. The Wilson line generates masses just as a Vacuum Expectation Value (VEV) in the corresponding representation would do and thus it is common to interpret the Wilson line as an effective VEV. It is important to remark, however, that even though one can use the same formalism and notation for the Wilson lines as if they were VEVs, they are not, i.e. they do not arise from the minimization of any potential, and effectively describe the mass splittings due to the different extra dimensional profiles arising from boundary conditions on translations. 
The fields that obtain a mass through the Wilson lines are thus not present in the low energy theory. The three independent Wilson lines are effectively described by VEVs for the SM singlet fields
\begin{equation} 
\braket{\phi_1}: \textcolor{brown}{\braket{\Phi}},\ \ \
\braket{\phi_2}: \textcolor{blue}{\braket{F^c}},\ \ \
\braket{\phi_3}:\textcolor{blue}{\braket{f^c}},
\end{equation}
where the $\textcolor{blue}{\braket{F^c}},\textcolor{blue}{\braket{f^c}}$ are aligned in the right handed sneutrino component. 

Note that these  effective VEVs in general does not preserve $\mathcal{D}$ flatness as
\begin{equation}
\braket{\mathcal{D}_A}=\sum_{i,j=1,}^3\sum_{a,b=2}^3\left[\braket{\textcolor{violet}{\nu^c_i}}^\dagger T_A \braket{\textcolor{violet}{\nu^c_j}}+\braket{\textcolor{brown}{\varphi_b}}^\dagger T_A \braket{\textcolor{brown}{\varphi_a}}+\braket{\textcolor{violet}{\nu^c_i}}^\dagger T_A\braket{\textcolor{brown}{\varphi_a}} +\braket{\textcolor{brown}{\varphi_a}}^\dagger T_A \braket{\textcolor{violet}{\nu^c_i}}\right],
\end{equation}
does not necessarily vanish, therefore breaking the remaining SUSY.

This describes the complete orbifold breaking process at the compactification scale that leads to a low energy model with the SM gauge group $SU(3)_C\times SU(2)_L\times U(1)_Y$ with the following fields:
\begin{eqnarray}\nonumber
V_\mu &:&  \textcolor{magenta}{(\textbf{3},\textbf{1},0)}+\textcolor{magenta}{(\textbf{1},\textbf{3},0)}+
	\textcolor{magenta}{(\textbf{1},\textbf{1},0)},
	\\ 
\phi_1 &:& \textcolor{green}{(\textbf{1},\textbf{2},3)}+\textcolor{green}{(\textbf{1},\textbf{2},-3)}+ 2\times
	\textcolor{brown}{(\textbf{1},\textbf{1},0)}+ 2\times (\textbf{3},\textbf{1},-2)+2\times (\bar{\textbf{3}},\textbf{1},2),
	\\ \nonumber
\phi_2 &:& \textcolor{blue}{(\textbf{3},\textbf{2},1)}+ \textcolor{blue}{(\textbf{1},\textbf{2},-3)}+2\times
	\textcolor{blue}{(\textbf{1},\textbf{1},6)} 
	 + 2\times\textcolor{blue}{(\bar{\textbf{3}},\textbf{1},-4)}+2\times\textcolor{blue}{(\bar{\textbf{3}},
	 \textbf{1},2)}
	+2\times \textcolor{violet}{(\textbf{1},\textbf{1},0)},
	\\ \nonumber 
\phi_3 &:&  2\times\textcolor{blue}{(\textbf{3},\textbf{2},1)}+ 2\times\textcolor{blue}{(\textbf{1},
	\textbf{2},-3)}+\textcolor{blue}{(\textbf{1},\textbf{1},6)} 
 	+ \textcolor{blue}{(\bar{\textbf{3}},\textbf{1},-4)}+\textcolor{blue}{(\bar{\textbf{3}},\textbf{1},2)}
	+ \textcolor{violet}{(\textbf{1},\textbf{1},0)},
\label{eq:zmff}
\end{eqnarray}
which can be named as
\begin{eqnarray}
\label{eq:mssm-plus-content}
\nonumber
V_\mu &:&  \textcolor{magenta}{G_\mu}+\textcolor{magenta}{W_\mu}+\textcolor{magenta}{B_\mu},
\\ \nonumber
\phi_1&:&\textcolor{green}{h_{u}}+\textcolor{green}{h_{d}}+\textcolor{brown}{\varphi_a}+t_a+\bar{t}_a,
\\ \nonumber
\phi_2&:& \textcolor{blue}{Q_1}+\textcolor{blue}{L_1}+\textcolor{blue}{e^c_a}+\textcolor{blue}{u^c_a}+\textcolor{blue}{d^c_a}+ \textcolor{violet}{\nu^c_a}, 
\\
\phi_3&:&\textcolor{blue}{Q_a}+\textcolor{blue}{L_a}+\textcolor{blue}{e^c_1}+\textcolor{blue}{u^c_1}+\textcolor{blue}{d^c_1}+ \textcolor{violet}{\nu^c_1},
\end{eqnarray}
with $a=2,3$. This is the MSSM field content plus three right handed neutrinos, two flavon singlets, and two vector-like triplets.

As mentioned above, this letter does not present any phenomenological results. However there are some interesting observations that can be noticed at this point. Note from Eq.~\eqref{eq:n41lag} that the only chiral field superpotential inherited from the 10d theory is 
\begin{equation} \label{eq:superpotential}
\begin{split}
\mathcal{W}(\phi_i)\sim \phi_1\phi_2\phi_3\sim& \textcolor{green}{h}\textcolor{blue}{F^c F}+ T\textcolor{blue}{F^c f^c}+T\textcolor{blue}{F f}
\\ =&\textcolor{green}{h_u}\textcolor{blue}{ u^c_a Q_a}+\textcolor{green}{h_u}\textcolor{violet}{\nu^c_a}\textcolor{blue}{L_a}+\textcolor{green}{h_d}\textcolor{blue}{ d^c_a Q_a}+\textcolor{green}{h_d}\textcolor{blue}{ e^c_a L_a}\\
&+t_a\textcolor{violet}{\nu^c_a}\textcolor{blue}{d^c_1}+t_a\textcolor{violet}{\nu^c_1}\textcolor{blue}{d^c_a}+\bar{t}_a\textcolor{blue}{Q_1}\textcolor{blue}{L_a}+\bar{t}_a\textcolor{blue}{Q_a}\textcolor{blue}{L_1}.
\end{split}
\end{equation}

The first four terms on the RHS (second line) of this equation are the usual Yukawa couplings. Note that at this level, once electroweak symmetry gets triggered by the VEVs of the higgses, the first-generation fermions are massless and the rest are universal with up- and down-type (neutrino and charged lepton) only differentiated by $\tan\beta$. This universality gets broken thanks to a large sneutrino VEV that gives (a large) mass to the triplets $t_a$ and $\bar{t}_a$ in $\phi_1$ in Eq.~\eqref{eq:mssm-plus-content} and mixes them with the down-type quarks, and also due to the R parity violating term (second term in Eq.~\eqref{eq:superpotential}). Masses for the first generation fermions will have to be generated radiatively (mediated by KK modes). In fact, the same must be true for the needed large Majorana masses of right-handed neutrinos and for the $\mu$ term, which are not present at tree level (If these effects can in fact work to generate a phenomenologically viable model is a work under progress. Note however that they would happen after SUSY breaking and, being radiative processes, they woould not affect the superpotential thanks to the non-renormalization theorem).

The last four terms in the equation contain the couplings of the fermions to the vector-like triplets. It is easy to see that integrating them out would generate effective couplings of the form $QQLL$ which do not produce proton decay.

 \section{Anomaly cancelation}
 
Any consistent QFT must be fundamentally anomaly free. 
This theory has an $E_8$ gauge symmetry 
and 10d bulk matter arising from the gauge supermultiplet $\mathcal{V}_{(\textbf{248})}\sim  \textbf{248}$, which is made of a 10d vector field and a 10d Majorana/Weyl fermion. The representation of the 10d chiral fermions is real, therefore the fundamental 10d theory is free of gauge anomalies. 
However, anomalies may arise in the compactified theory due to the discontinuous nature of the extra dimensions~\cite{Scrucca:2004jn}.

In order to explore the consequences of this, in the effective low energy theory,
consider an orbifold $T^6/(\mathbb{Z}_6\times \mathbb{Z}_2)$, where any bulk field can be decomposed as a Fourier like series of extra dimensional rotation $SO(6)$ modes, separated by orbifold transformation eigenstates 
\begin{equation}
\Psi(x,z_i)=\sum_{s=0}^\infty\sum_{n=0}^{5}\sum_{m=0}^{1} \psi_{n+6 s,m+ s}(x) f_{n+6 s,m+ s}(z_i),
\end{equation}
with $f_{n,m}$ as the orbifold eigenstates with
\begin{eqnarray} \nonumber
 R f_{n,m}(z_i)=\alpha^n (-1)^m f_{n,m}(z_i),
\end{eqnarray}
where $R$ is the representation of the $\mathbb{Z}_6\times \mathbb{Z}_2$ rotation acting on the superfield. 
Their explicit form is not required for this analysis, up to the fact that $f_{00}(z_i)=f_{00}$ is a constant. The $V_\mu,\phi_{1,2,3}$ all decompose as $\Psi$.
It is important to note that the original $(\textbf{248})$ representation is decomposed into $12$ different representations (one for each $(n,m)$ value) of the unbroken gauge group, each one with a different eigenvalue under the orbifold rotation. 

After compactification (integrating out the extra dimensions), a 10d field $\Psi(x,z_i) \sim \textbf{248}$ is represented as an infinite tower of 4d $\psi_{s}(x)$ ($s = 1 \dots \infty$) fields, each one completing a full $(\textbf{248})$ of $E_8$. Thus, at this stage the theory is still anomaly free. Now one must determine the low energy effective theory. Recall that, as mentioned above, the low energy theory will be composed of the massless modes obtained by compactification involving both rotation and translation boundary conditions. 
Thus, the complete set of zero -massless - modes in the low energy theory are obtained by identifying those associated to the rotation boundary conditions and subtracting those that get a mass through the Wilson lines.  The zero modes associated to the boundary conditions in the compactification are identified as usual, by finding the $f_{n,m}(z_i)$ eigenfunctions of the ED derivatives with $0$ eigenvalue. All others are massive modes and become arranged into vector-like pairs (to obtain a mass term).  A very interesting observation is that if a specific mode is in a representation where both itself and its complex conjugate lack zero modes, then they pair up inside the same $s$ set to form a mass term. If not, then the complex conjugated representation of the zero mode is paired with the one from $s+1$ set. This way all the non-zero modes are paired up into vector-like pairs, canceling all anomalies not associated to the zero modes. This would require the zero modes to be anomaly free by themselves, something that is not automatic in general. In the model presented in this letter one sees that the zero modes obtained by the full orbifold compactification, listed in Eq.~\eqref{eq:zmff}, are anomaly free for the remaining gauge symmetry.

As a final check, one must consider that in orbifolds there are singular points where additional anomalies can arise~\cite{vonGersdorff:2006nt}. 

 As the scenario presented in this letter only contains a single bulk multiplet, there are no localized fields in the fixed points nor in the fixed tori.

There are no chiral representations in the fixed tori, therefore there are no anomalies generated there. At the fixed points some fields always vanish and therefore they lack the full gauge symmetry. The anomaly must be studied at each fixed point with their corresponding gauge symmetry and field content.

The fixed points, due to their discontinuous nature, can only be studied consistently in the gauge where the Wilson lines are seen as an effective VEV, namely in the low energy unbroken gauge symmetry (in other choices of gauge the wavefunction is multiply defined).

 The orbifold used in this construction has the following fixed points (with $\alpha = e^{i\pi/3}$)
\begin{equation}
\bar{z}_{4d}=\left(x,\left\{0,\frac{1+\alpha}{3},\frac{2+2\alpha}{3}\right\},\left\{0,\frac{1}{3},\frac{2}{3}\right\},\left\{0,\frac{1}{3},\frac{2}{3}\right\}\right).
\end{equation}
Each transformation will leave some points invariant, and will permute the others. After applying all transformations, these points can only have the SM gauge symmetry with representations in Eq.~\eqref{eq:zmff} as field content. Therefore they are anomaly free. 

It should be emphasized that gravitational effects lie beyond the scope of this letter. As there is no consistent QFT of gravity yet, its consistency with this model can't be fully addressed.
 
Note that if this was string theory (or a string motivated model), the string would decompose into KK modes plus winding modes. They both would share the same zero mode, thus not forming complete $(\textbf{248})$ representations. Furthermore, in string theory there are inherently localized states which only feel a fraction (or none) of the Wilson lines, so the continuous Wilson lines can't play a role in anomaly cancelation, as they do in a QFT with only bulk states \cite{Scrucca:2004jn}. 

It has been shown that the model is free of gauge anomalies at all levels, however there are Poincar\`e anomalies. These arise as the gaugino is a 10d chiral (left by definition) fermion and since this is the only fermion  (in the ($\textbf{248}$) representation), it generates a global Lorentz anomaly. It becomes relevant once gravity is considered. One can add gravity by upgrading to local the global Poincar\`e symmetry, which is equivalent to full coordinate invariance, i. e. General Relativity. The simplest (but not unique) way to cancel the anomaly is to add an $E_8$ ($\textbf{248}$) 10d right chiral supermultiplet or 248 $E_8$ singlets. If instead one localizes the global SUSY then there is also a spin $3/2$ 10d chiral gravitino. The minimal way to cancel the anomaly is to add 247 singlet 10d left chiral superfields, although it is more symmetrical to add a single 10d right chiral superfield $E_8$ singlet (usually called dilaton) and an $E_8$ ($\textbf{248}$) 10d left chiral supermultiplet \cite{Green:1984sg}. As in string theory an adjoint chiral multiplet can't be added, one must add a second gauge symmetry $E_8$. Finally gravity can be an emergent phenomena instead of a fundamental one \cite{deAnda:2019tri,Linnemann:2017hdo,Barcelo:2001ah}. In this case the 10d SupePoincar\`e symmetry stays global and its anomaly is not a problem (it is broken by the orbifold compactification anyway). If the emergent local symmetry arises after compactification one would have effective 4d gravity. As the field content in this model does not generate any 4d SuperPoincar\`e anomaly, no extra field content would be required. 
As none of the discussed QFTs of gravity has deemed to be fully consistent and each setup requires different constraints, gravity lies beyond the scope of this model.

 \section{Conclusion}
The possibility that all interactions present at low energies and described by the SM gauge structure are {\it united} into a single gauge group at high energies is both beautiful and old. The objective that led to this letter consisted in answering the question of whether or not it is possible, within the QFT framework alone, to have a pure gauge - {\it complete} - $E_8$ unification model consistent with the SM (particle content and gauge group). The response given here is that a complete $E_8$ unified model in $10$d, composed of the single vector superfield of the gauge symmetry, compactified on the orbifold $T^6/(\mathbb{Z}_6\times \mathbb{Z}_2)$ including its associated Wilson lines, leads to a low energy model with the field content (given in Eq.~\eqref{eq:zmff}) that corresponds to the MSSM one plus three right handed neutrinos, two singlets, and two vector-like triplet pairs. Thus, this set up is anomaly free under the remaining $SU(3)_C\times SU(2)_L\times U(1)_Y$ SM gauge group at all energies below the compactification scale, as desired. While this has not been shown to be a phenomenologically viable model, it proves that full unification of the SM field content into a single multiplet is possible in QFT without the need of any extra field content. This has been a widely ignored as it is not possible in string theory. This letter does not address gravity. An investigation of the phenomenology is currently being pursued, however a few remarks were presented that motivate its further study.

\bibliographystyle{ieeetr}

\begin{thebibliography}{1}


\bibitem{Georgi:1974sy}
H.~Georgi and S.~L.~Glashow,
Phys. Rev. Lett. \textbf{32} (1974), 438-441
doi:10.1103/PhysRevLett.32.438

\bibitem{Fritzsch:1974nn}
H.~Fritzsch and P.~Minkowski,
Annals Phys. \textbf{93} (1975), 193-266
doi:10.1016/0003-4916(75)90211-0

\bibitem{Mohapatra:1974gc}
R.~N.~Mohapatra and J.~C.~Pati,
Phys. Rev. D \textbf{11} (1975), 2558
doi:10.1103/PhysRevD.11.2558

\bibitem{Pati:1974yy}
J.~C.~Pati and A.~Salam,
Phys. Rev. D \textbf{10} (1974), 275-289
doi:10.1103/PhysRevD.10.275

\bibitem{Langacker:1980js}
P.~Langacker,
Phys. Rept. \textbf{72}, 185 (1981)
doi:10.1016/0370-1573(81)90059-4

\bibitem{Tanabashi:2018oca}
M.~Tanabashi \textit{et al.} [Particle Data Group],
Phys. Rev. D \textbf{98}, no.3, 030001 (2018)
doi:10.1103/PhysRevD.98.030001

\bibitem{Nilles:2014owa}
H.~P.~Nilles and P.~K.~S.~Vaudrevange,
Mod. Phys. Lett. A \textbf{30} (2015) no.10, 1530008
doi:10.1142/S0217732315300086
[arXiv:1403.1597 [hep-th]].


\bibitem{Slansky:1981yr}
  R.~Slansky,
  Phys.\ Rept.\  {\bf 79} (1981) 1.
  \doi{10.1016/0370-1573(81)90092-2}

\bibitem{Adler:2002yg}
  S.~L.~Adler,
  Phys.\ Lett.\ B {\bf 533} (2002) 121
  \doi{10.1016/S0370-2693(02)01596-4}
  [hep-ph/0201009].

\bibitem{Aranda:2020noz}
A.~Aranda, F.~J.~de Anda and S.~F.~King,
[arXiv:2005.03048 [hep-ph]].

\bibitem{Olive:1982ai}
D.~I.~Olive and P.~C.~West,
Nucl. Phys. B \textbf{217} (1983), 248-284
doi:10.1016/0550-3213(83)90086-X

\bibitem{Babu:2002ti}
  K.~S.~Babu, S.~M.~Barr and B.~Kyae,
  Phys.\ Rev.\ D {\bf 65} (2002) 115008
  \doi{10.1103/PhysRevD.65.115008}
  [hep-ph/0202178].

\bibitem{ArkaniHamed:2001tb}
N.~Arkani-Hamed, T.~Gregoire and J.~G.~Wacker,
JHEP \textbf{03} (2002), 055
doi:10.1088/1126-6708/2002/03/055
[arXiv:hep-th/0101233 [hep-th]].

\bibitem{Brink:1976bc}
L.~Brink, J.~H.~Schwarz and J.~Scherk,
Nucl. Phys. B \textbf{121} (1977), 77-92
doi:10.1016/0550-3213(77)90328-5

\bibitem{Fischer:2012qj}
M.~Fischer, M.~Ratz, J.~Torrado and P.~K.~S.~Vaudrevange,
JHEP \textbf{01} (2013), 084
doi:10.1007/JHEP01(2013)084
[arXiv:1209.3906 [hep-th]].

\bibitem{Scrucca:2004jn}
C.~A.~Scrucca and M.~Serone,
Int. J. Mod. Phys. A \textbf{19} (2004), 2579-2642
doi:10.1142/S0217751X04018518
[arXiv:hep-th/0403163 [hep-th]].

\bibitem{vonGersdorff:2006nt}
G.~von Gersdorff,
JHEP \textbf{03}, 083 (2007)
doi:10.1088/1126-6708/2007/03/083
[arXiv:hep-th/0612212 [hep-th]].

\bibitem{Green:1984sg}
M.~B.~Green and J.~H.~Schwarz,
Phys. Lett. B \textbf{149} (1984), 117-122
doi:10.1016/0370-2693(84)91565-X

\bibitem{deAnda:2019tri}
F.~J.~De Anda,
Class. Quant. Grav. \textbf{37} (2020) no.19, 195012
doi:10.1088/1361-6382/aba31a
[arXiv:1910.03599 [hep-th]].

\bibitem{Linnemann:2017hdo}
N.~S.~Linnemann and M.~R.~Visser,
Stud. Hist. Phil. Sci. B \textbf{64} (2018), 1-13
doi:10.1016/j.shpsb.2018.04.001
[arXiv:1711.10503 [physics.hist-ph]].

\bibitem{Barcelo:2001ah}
C.~Barcelo, S.~Liberati and M.~Visser,
Class. Quant. Grav. \textbf{18} (2001), 3595-3610
doi:10.1088/0264-9381/18/17/313
[arXiv:gr-qc/0104001 [gr-qc]].
\end{thebibliography}

\end{document}